\documentclass[12pt]{iopart}

\usepackage{iopams}  
\usepackage{graphicx}
\usepackage{xcolor}
\usepackage{wasysym}
\usepackage{amsbsy}
\usepackage{gensymb}
\begin{document}

\title{Quasistatic approximation in neuromodulation}

\author{Boshuo Wang$^1$, Angel V. Peterchev$^{1,2,3,4}$, Gabriel Gaugain$^5$, Risto J. Ilmoniemi$^6$, Warren M. Grill$^{2,3,4,7}$, Marom Bikson$^8$, and~Denys~Nikolayev$^{5,*}$}
\address{$^1$ Department of Psychiatry and Behavioral Sciences, Duke University, Durham, NC 27710, USA\\
$^2$ Department of Electrical and Computer Engineering, Duke University, Durham, NC 27708, USA\\
$^3$ Department of Biomedical Engineering, Duke University, Durham, NC 27708, USA\\
$^4$ Department of Neurosurgery, Duke University, Durham, NC 27710, USA\\
$^5$ Institut d’électronique et des technologies du numérique (IETR UMR 6164), CNRS~/~University of Rennes, 35000 Rennes, France\\
$^6$ Department of Neuroscience and Biomedical Engineering, Aalto University School of Science, Espoo, Finland\\
$^7$ Department of Neurobiology, Duke University, Durham, NC 27710, USA\\
$^8$ The City College of New York, New York, NY 11238, USA\\
$^*$ Author to whom any correspondence should be addressed.}
\ead{\mailto{denys.nikolayev@deniq.com}}
\vspace{10pt}
\begin{indented}
\item[]January 2024
\end{indented}

\begin{abstract}
We define and explain the quasistatic approximation (QSA) as applied to field modeling for electrical and magnetic stimulation. Neuromodulation analysis pipelines include discrete stages, and QSA is applied specifically when calculating the electric and magnetic fields generated in tissues by a given stimulation dose. QSA simplifies the modeling equations to support tractable analysis, enhanced understanding, and computational efficiency. The application of QSA in neuro-modulation is based on four underlying assumptions: (A1) no wave propagation or self-induction in tissue, (A2) linear tissue properties, (A3) purely resistive tissue, and (A4) non-dispersive tissue. As a consequence of these assumptions, each tissue is assigned a fixed conductivity, and the simplified equations (e.g., Laplace’s equation) are solved for the spatial distribution of the field, which is separated from the field’s temporal waveform. Recognizing that electrical tissue properties may be more complex, we explain how QSA can be embedded in parallel or iterative pipelines to model frequency dependence or nonlinearity of conductivity. We survey the history and validity of QSA across specific applications, such as microstimulation, deep brain stimulation, spinal cord stimulation, transcranial electrical stimulation, and transcranial magnetic stimulation. The precise definition and explanation of QSA in neuromodulation are essential for rigor when using QSA models or testing their limits.
\end{abstract}

\vspace{2pc}

\clearpage


\section{Introduction}

Computational modeling of electromagnetic fields has broad application in neuromodulation to estimate delivered intensity, predict and optimize response, and inform risk. The quasistatic approximation (QSA)~\cite{Plonsey1967Considerations, Lindsay2004From,Bossetti2008Analysis,Makarov2018Quasi-Static,Caussade2023Towards,Gaugain2023Quasi-static,Unal2023Quasi-static}, applied under the associated quasistatic assumptions, is ubiquitous in computational modeling of electrical and magnetic stimulation. The purpose of QSA is to reduce the computational complexity of calculating the electric fields (E-fields), potentials, and currents in the target tissue, e.g., brain, spinal cord, or peripheral nerve. Despite its prevalence in neuromodulation modeling, QSA as used in neuromodulation has not been clearly defined, and the underlying assumptions are typically implicit or, at times, ambiguous. Thus, the use of QSA, as understood and applied in neuromodulation modeling, needs to be defined. Here we explain how QSA is applied in a consistent manner across modeling of neuromodulation under four underlying assumptions, and we derive the corresponding governing equations. We also review adaptations of the QSA simulation pipeline to accommodate cases where some of the simplifying assumptions do not hold. We further provide historical and application-specific examples and clarify QSA-related terminology as applied in general electromagnetism. Finally, we provide recommendations for communicating the use of QSA in modeling studies.

\subsection{Scope}

The scope of this paper includes the computational methods used to predict the delivery of electromagnetic energy to the body during neuromodulation---and specifically the application of QSA. 
These computations yield the E-fields in the tissue of interest, and/or related quantities including tissue potentials, current densities, or magnetic fields.
The solutions provide estimates of these variables, which are subject to model assumptions and verification.
We define these assumptions and related parameters, focusing on the E-field, as the magnetic field is not influenced by the tissues when QSA is valid, and as the E-field is understood to mediate the effects of the most common forms of neural stimulation \cite{Peterchev2012Fundamentals}. Nonetheless, the magnetic field has a key role in neural stimulation through electromagnetic induction, and QSA typically applies to the magnetic field as well; therefore, we occasionally note generalization to the magnetic field. 
To frame the use of QSA, we also broadly consider additional steps in modeling pipelines, before or following the prediction of the E-field.
Quasistatic analysis of approaches for recording bioelectric phenomena, such as the electrocardiogram and impedance plethysmography, was summarized by Plonsey and Heppner  \cite{Plonsey1967Considerations}. 
Specific to the nervous system, QSA has been applied to analyses of endogenous brain activity as recorded by electroencephalography (EEG) \cite{Stinstra1998volume,Bédard2004Modeling} and magnetoencephalography (MEG) \cite{Stinstra1998volume,Hämäläinen1993Magnetoencephalographytheory}.
Subsequently, QSA in neuromodulation was adapted from these analyses of bioelectric signal recording. 
In contrast to endogenous electromagnetic fields, however, neuromodulation uses exogenous fields and applies a wide range of stimulation waveforms with much broader frequency content and much higher intensities than intrinsic signals, and thus warrants distinct considerations.
The application of QSA in modeling neuromodulation encompasses several modalities, including:

\begin{enumerate}
    \item Non-invasive electrical and magnetic stimulation techniques:
    \begin{enumerate}
        \item Transcranial electrical stimulation (tES), including transcranial direct current, alternating current, random noise, and pulsed current stimulation (tDCS, tACS, tRNS, and tPCS, respectively), temporal interference stimulation (TIS), intersectional short pulse stimulation (ISPS), cranial electrotherapy stimulation (CES), pulsed suprathreshold transcranial electrical stimulation (TES), and electroconvulsive therapy (ECT).
        \item Transcutaneous electrical nerve stimulation (TENS), including peripheral and cranial nerves, such as transcutaneous (auricular) vagus nerve stimulation (tVNS/taVNS).
        \item Transcutaneous spinal cord stimulation (tSCS).
        \item Transcranial magnetic stimulation (TMS), and variants including low-field magnetic stimulation (LFMS), pulsed electromagnetic field therapy (PEMFT), and magnetic seizure therapy (MST).
    \end{enumerate}
    \item Invasive stimulation methods:
    \begin{enumerate}
        \item Cortical stimulation.
        \item Intracortical microstimulation (ICMS).
        \item Deep brain stimulation (DBS).
        \item Vagus nerve stimulation (VNS).
        \item Spinal cord stimulation (SCS) and dorsal root ganglion (DRG) stimulation.
        \item Peripheral nerve stimulation (PNS).
        \item Cochlear implants, electrical retinal/visual prostheses, and motor and sensory prostheses using implanted electrodes or microelectrode arrays (MEA).
        \item Intracortical magnetic stimulation with microcoils.
    \end{enumerate}
\end{enumerate}

We note that tDCS has a nominally constant amplitude, and thus, it might not require the “quasi” qualifier. 
However, since tissue properties and current amplitude may change over time, certain QSA assumptions may be needed. 
Further, the term ``quasistatic'' is limited in the neuromodulation literature to electric and magnetic stimulation with signals having frequency content below approximately 100\,kHz. 
Consequently, modeling approaches based on QSA do not encompass higher-frequency electromagnetic modulation techniques in human-sized targets in the\,MHz to\,GHz range \cite{Apollonio2013Feasibility,Oh20216.5-GHz,Ahsan2022EMvelop} or light-based stimulation using infrared light or optogenetics \cite{Williams2015Optogenetic, Wang2022Transcranial}. 
Additionally, neuromodulation with ultrasound \cite{Tufail2011Ultrasonic,Kamimura2020Ultrasound,Legon2022Transcranial}, a wave-based phenomenon that is not governed by Maxwell’s equations, falls outside the domain of our discussion.

\subsection{Motivation}

QSA as applied in neuromodulation modeling is clearly not strictly “static”---where time is not a factor. 
Rather, QSA considers conditions where the physics of neuromodulation can be represented using \textit{simplifications} that in a specific context are static-\textit{like}---hence “approximation” and “quasi”. 
As detailed below, central to QSA in neuromodulation is the separability of the spatial and temporal components of the electromagnetic fields, so that static equations (e.g., Laplace’s equation) and solvers can be used to obtain the spatial distribution of the E-field. 
Such simplification is both more interpretable and less computationally intensive compared to solving the complete set of Maxwell's equations.

Similar to all mathematical simplifications, the motivation for QSA is to 1)~reduce computational complexity and thus the required resources for calculations and simulations; 2)~support explanatory models and design-oriented analysis \cite{Middlebrook1992Methods}, which always require an appropriate level of approximation; and 3)~minimize ambiguity of model parameters such as tissue electrical properties. 
In most practical cases, QSA is a necessity, given the complexity of the alternative, and its use has become, in large part, driven by expediency. 
QSA is so ubiquitous and ingrained in neuromodulation field modeling that its underlying conditions and implications are mostly implicit or unrecognized.

The purpose of this review is to define QSA as used in neuromodulation and explicitly describe its assumptions and integration into overall modeling pipelines. 
While QSA and its assumptions are rarely fully or consistently explained (if at all) when the term is used in the literature, we do not suggest that there is a specific deficiency or misuse in the field. 
Rather, there is a broad implicit consensus across neuromodulation domains and decades of experience with diverse techniques on how to apply QSA. 
Our overarching aim is to make clear the underlying assumptions---which we classify in four categories---as applied across the neuromodulation literature (section 2).
This review is “descriptive” of existing practice rather than “prescriptive” of any required changes. 

Irrespective of its history and validation, we do not aim to endorse or criticize the use of QSA in neuromodulation or quantify the errors arising from the approximation (which are application specific, see section 3). 
Nonetheless, we conclude this paper with recommendations to promote transparency and reproducibility in the communication of QSA in neuromodulation. 
Rational use of computational models of neuromodulation---including sensitivity analyses of modeling parameters, testing the validity of assumptions, and developments of new techniques---requires explicit recognition of the QSA based on the four assumptions.

\subsection{Organization}

Section 2 presents the principles and applications of QSA in neuromodulation. 
We first describe the multistage modeling framework widely used in neuromodulation and the specific role of QSA (section 2.1). 
Four underlying assumptions of QSA (heretofore common but implicit) are defined and explained (section 2.2). We then provide the equations governing QSA (section 2.3) as a consequence of these assumptions. 
The frequency spectra of common stimulation pulses and frequency dependency of tissue electrical parameters are examined under the consideration of QSA (section 2.4). 
We explain the separability of space and time under QSA (section 2.5), which underpins its application in modeling practice. 
Adaptations of QSA, even under conditions where the four assumptions are understood to not strictly hold (e.g., dispersive, capacitive, or nonlinear tissue) are described (section 2.6). 
We also provide a perspective on terminology in general computational electromagnetics, to disambiguate these terms from QSA in neuromodulation (section 2.7). 
In section 3, we delve into the historical adaptation of QSA, first in bioelectricity, then neuromodulation in general (sections 3.1 and 3.2), and then in specific applications (sections 3.3–-3.6). 
Here, we summarize validation efforts and prior analyses of limitations of QSA.
In section 4, we present recommendations for improved reporting of QSA applications in neuromodulation modeling papers, based around the four underlying assumptions. 

\section{Definitions and implications}

\subsection{Multi-stage framework for modeling of neuromodulation}

\begin{figure*}[t]
	\centering
		\includegraphics[width = \textwidth]{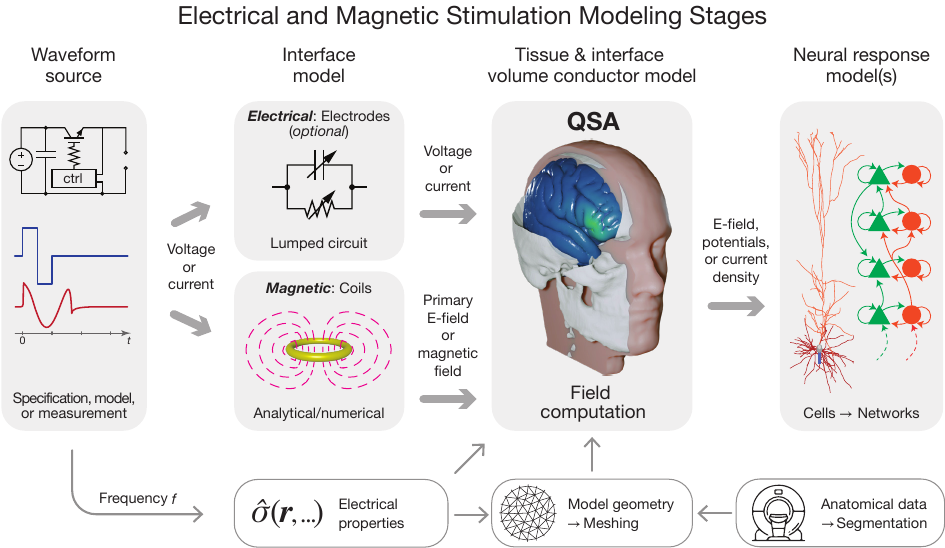}
	  \caption{\textbf{Conventional neuromodulation modeling pipelines with sequential steps including calculation of tissue E-fields}. Notwithstanding diversity in neuromodulation technologies and modeling approaches, modeling pipelines generally employ a sequential approach centered around the prediction of E-fields in the tissue---it is specifically at this step that QSA is applied. Steps preceding E-field simulation include waveform specification (whether defined, modeled, or measured) and simulation of coil/electrode output, and steps following E-field simulation involve producing neurophysiological, network, or behavior consequences. These steps may vary or be omitted depending on the modeling pipeline. Segmentation of tissues to define the model geometry and assignment of tissue electrical properties are essential in E-field modeling; the assignment of electrical properties involves QSA assumptions. Under QSA, the E-field calculation involves only the field spatial distribution, as time is separate at this specific stage. Not shown in this pipeline are numerous additional processing steps both in support of E-field modeling and broader integration into dose delivery and optimization, but the step of calculating the E-field, where QSA is applied, remains sequestered and key.}\label{fig1} 
\end{figure*}

Modeling of neuromodulation involves multiple distinct stages (Fig.~\ref{fig1}), including, but not limited to, stimulation device operation (electronics and transducers), tissue segmentation (image processing) and geometric model definition, electromagnetic field and current flow simulation in the biological tissue, estimation of tissue responses (electrical, thermal, and mechanical), prediction of neurophysiological and behavioral consequences, and any further integration of the model into a system (e.g., closed loop).These stages typically comprise separate sequential simulations or analyses. 
The scope of modeling as discussed in this paper, specifically as it relates to QSA in neuromodulation, applies to the calculation of the E-field within the tissue of interest.

Any stage of modeling equipment (the voltage or current source) is prior to the stage of calculating tissue E-field. 
As an independent stage, it is not subject to the same constraints and assumptions as other stages. 
The output of this stage is the current or voltage waveform applied to the subsequent stage. 
In most pipelines, the waveform is simply specified by parameters (e.g., device waveform settings) or obtained from recordings.

In some modeling pipelines, components of the stimulation device (electrodes or coils) are combined with the tissue when the E-field is computed. 
This combination preserves QSA as long as all underlying assumptions applied to the tissue also apply to the device components.
Alternatively, the influence of the electrodes and their interface with tissue can be modeled in a step after defining the current or voltage source and preceding tissue E-field calculation \cite{Wei2009Impedance,Lempka2018Characterization}, in which case the electrodes are not subject to QSA.
For example, a lumped-circuit model can represent the impedance of the electrode-tissue interface.
Similarly, the magnetic field generated by a TMS coil can be computed prior to simulation of the induced E-field in tissue, and can account, for example, for nonlinear effects due to saturation of a ferromagnetic core in the coil \cite{Makaroff2023Modeling}. 
These approaches preserve QSA for the E-field in the tissue, allowing for modeling adaptations (section 2.6) to deal with sources with frequency-dependent or nonlinear effects, such as charge transfer conductances of the electrode–electrolyte interface in electrical stimulation and skin effect in coil windings or saturation of ferromagnetic core materials in magnetic stimulation.

For the E-field modeling stage, the spatial sampling resolution of the tissue is typically on a macroscopic scale, i.e., larger than its underlying cellular composition \cite{Bédard2011Generalized}. 
Consistent with this scale, the tissue’s electromagnetic parameters are assumed to be aggregate bulk “averages” without considering the microscopic structures and nonlinear properties of individual cells. 
When field modeling is conducted with microscopic scales explicitly presented, QSA can be similarly used, provided that the relevant elements (such as membrane, intra- and extracellular media) are considered resistive \cite{Pucihar2006Numerical,Khadka2022Neurocapillary-Modulation,Weise2022effect,Khadka2023Multi-scale}.
Such meso- or microscopic models can predict macroscale tissue properties, e.g., with axons contributing to conductivity anisotropy and meninges reducing the effective conductivity of the cerebrospinal fluid space \cite{Pelot2019On,Jiang2020Enhanced}.

The temporal dynamics of the E-field are separated from its spatial distribution under QSA in the field simulation stage of modeling, and these dynamics are reintroduced in the subsequent stages to predict the neuronal response to stimulation \cite{Joucla2014Current,Ye2022Neuron}. 
The field distribution is applied to multi-compartment neuron models as extracellular potentials (or quasipotentials for magnetic stimulation \cite{Wang2018Coupling})---using either the quasi-uniform field assumption (typical for non-invasive stimulation) \cite{Bikson2013“Quasi-Uniform”,Khadka2019Quasi-uniform} or sampled on the microscopic scale of the neuronal morphology (typical for invasive stimulation).
The E-field is scaled by the temporal waveform, and the neuronal response is then simulated using conductance-based cable models hosting time-dependent nonlinear membrane dynamics (e.g., Hodgkin–Huxley-type ion channels). 
Typically, the model predicts the hyper- or depolarization of neuronal membranes and/or neural activation threshold \cite{Warman1992Modeling,McIntyre1999Excitation,Joucla2014Current,Aberra2020Simulation,Esmaeilpour2021Temporal}, although other biophysical responses to the E-field such as aggregate neural states, tissue heating, or interstitial water flux may also be simulated \cite{Datta2009Bio-heat,Zannou2021Tissue,Khadka2023Multi-scale}. 
These predicted neuromodulation outcomes are time- and waveform-dependent.
Consequently, QSA and the term “quasistatic” in the neuromodulation modeling literature refer specifically to the E-field calculation stage of modeling.

\subsection{Underlying assumptions and implications of QSA in neuromodulation}

QSA in neuromodulation modeling is a set of approximations based on specific assumptions, applied specifically to the stage of E-field calculation in tissues. 
While the suitability of QSA may vary depending on the specific application, the underlying assumptions are generally accepted though implicit conventions and broad consensus within the neuromodulation field.
We define the following four assumptions on which QSA is based:

\begin{description}
    \item \textbf{A1}.~No wave propagation or self-induction in tissue
    \item \textbf{A2}.~Linear tissue properties
    \item \textbf{A3}.~Purely resistive tissue
    \item \textbf{A4}.~Non-dispersive tissue
\end{description}

Assumption A1 focuses on the relationship between the electric and magnetic fields, whereas assumptions A2--A4 mostly concern the tissue electrical parameters and the relationship between the E-field and the current density in the tissue (i.e., tissue impedance). 
The magnetic properties of biological tissue implicitly follow assumptions A2 and A4 by generally setting the relative permeability to unity, since the tissue permeabilities have negligible influence on the magnetic field \cite{Tharayil2018Field}. 
In the following sections, we describe the assumptions in detail and discuss their implication, validity, and relaxation.

\subsubsection{No wave propagation or self-induction in tissue (A1)\\}

Assumption A1 neglects the time derivatives relating the electric and magnetic fields when solving for them. 
Therefore, electromagnetic wave propagation and the resultant spatial attenuation (i.e., skin effect) and phase variation of the fields are assumed to be negligible and ignored in the field calculation.
This assumption is central to the quasistatic modeling approach, and, except when specifically analyzing and validating QSA in neuromodulation, this assumption is not relaxed.

For this assumption to be valid, the dimensions of the relevant tissue volume need to be much smaller than its electromagnetic skin depth to neglect attenuation and also much smaller than the wavelength of the electromagnetic field to neglect spatial phase variations. 
Equivalently for the latter, the time scale of interest is much larger than the wave propagation delay across the tissue \cite{Haus1989Electromagnetic,Jackson1999Classical}.

From the perspective of Maxwell's equations, neglecting the time derivatives also ignores the E-field induced by the magnetic field of the tissue current according to Faraday’s law and Lenz’s law (i.e., self-induction), which is sometimes listed separately as the “no induction” assumption  \cite{Plonsey1967Considerations,Bossetti2008Analysis,Caussade2023Towards,Gaugain2023Quasi-static}. 
For electrical stimulation, the current from external sources in the electrodes and leads is continuous and equivalent to the tissue current, so the inductive effects of the external sources are also negligible.
However, for magnetic stimulation (e.g., TMS), induction is essential: The E-field in the brain is proportional to the time derivative of the magnetic field generated by the TMS coil (i.e., mutual induction).
Despite the time derivative relationship, both the magnetic field and the induced E-field can be analyzed using quasistatic (more specifically, magneto-quasistatic) equations, as further discussed in sections 2.3.3 and 2.7.

\subsubsection{Linear tissue properties (A2)\\}

Assumption A2 states that the constitutive equations describing the electromagnetic fields in the tissue are linear \cite{Bédard2011Generalized}, and tissue properties (permittivity, permeability, and conductivity) do not depend on the amplitude of the local electromagnetic field, current flow, or other factors such as temperature \cite{Gholami-Boroujeny2015Theoretical,Wang2024Optimizing} that change in response to stimulation.

In the QSA literature, linearity is mentioned briefly, if at all, preceding detailed analysis of QSA itself \cite{Plonsey1967Considerations,Caussade2023Towards,Gaugain2023Quasi-static}. 
While seemingly trivial and unrelated to time, linearity implies time-invariance and therefore allows the computational complexity to be substantially decreased using frequency domain analysis, which requires a linear time-invariant system.
QSA studies with “time-harmonic” representation or frequency-dependent tissue properties assume linearity \cite{Butson2005Tissue,Bossetti2008Analysis,Grant2010Effect}.

The linearity assumption allows the E-field amplitudes to be calculated for only one stimulation intensity and simply scaled to obtain the solution for other amplitudes. 
Further, the solution to multiple stimulation sources can be computed as a linear superposition of the solutions for individual channels.
This is useful in applications such as high definition tES or patterned arrangement of cathodes and anodes on multi-electrode arrays, and it enables optimization of the electrode currents to achieve a specific E-field distribution \cite{Dmochowski2011Optimized,Peña2017Particle}.
In this context, appropriate boundary conditions should be used on the inactive electrodes \cite{Pelot2018Modeling}.
Similarly, linear superposition can be used to shape the field of magnetic stimulation with coil arrays \cite{Ruohonen1998Focusing,Koponen2018Multi-locus}. 
The linearity assumption is also a necessary condition (beside assumption A1) that allows the spatial and temporal components of the field to be separated, as discussed in sections 2.5 and 2.6. 
In neuromodulation modeling studies, the linearity assumption has been relaxed only in very limited cases, and so far, not together with other assumptions.

\subsubsection{Purely resistive tissue (A3)\\}
Under assumption A3, the time derivative relating the current density and E-field is neglected.
Only the free current density and thus the real part of the complex conductivity is considered. 
This removes the frequency- and location-dependent temporal phase shift between the E-field and current density distributions due to the tissue capacitance. 
This assumption is typically valid for dc or low permittivity media at low frequencies, and is commonly denoted as “no capacitance” or a “purely resistive” medium in the neuromodulation modeling literature.

Naturally following this assumption, the continuity condition at tissue boundaries for the current density (specifically its normal component) becomes the same as in the static ohmic case \cite{Plonsey1967Considerations,Grant2010Effect}. Therefore, the normal component of the tissue E-field is zero at external boundaries interfacing non-conductive media and there is only tangential E-field and current flow. 
In TMS, the E-field is approximately (in the spherical model, exactly) tangential also further away from the scalp \cite{Nummenmaa2013Comparison}.

The assumption of purely resistive tissue must be relaxed when evaluating the influence of tissue capacitance on the E-field and waveform. In such cases (see section 2.6), the complex conductivity has frequency dependence due to its capacitive component, regardless of assumption A4 (i.e., whether conductivity and permittivity parameters are themselves dispersive or not).

\subsubsection{Non-dispersive tissue (A4)\\}

Assumption A4 considers tissue properties to be non-dispersive, indicating that they remain frequency-independent, at least within the spectrum of the stimulation waveform. 
In practice, neuromodulation modeling studies typically make this assumption implicitly, along with assumptions A2 and A3, by assigning a single conductivity value to each tissue.
For stimulation using dc or a single frequency (e.g., tDCS or tACS, respectively), these tissue parameters reflect the magnitude of the impedance at the given frequency. 
In these cases, assumption A4 is not so much an assumption of frequency-independent tissue properties as the (implicit) recognition that the relevant properties correspond to a specific frequency. 
For the majority of cases (e.g., pulsed stimulation), waveforms are composed of multiple frequencies (see section 2.4 and Fig.~\ref{fig2}~A), and the use of a single parameter for each tissue reflects an approximation. 
Such an approximation may be justified by considering the central frequency content of the waveform, or otherwise sufficient given the goals of the stimulation study.
In all cases, the selection of single fixed purely-resistive tissue values (which involves assumptions A1, A2, and A3) should be recognized as a key step in model development (see section 4).

Despite not being explicitly considered by Plonsey and Heppner \cite{Plonsey1967Considerations} in their QSA analysis, the non-dispersion assumption (A4) is compatible with the other assumptions and broadens the space–time separability from a single frequency to arbitrary waveforms.
In neuromodulation modeling studies, assumption A4 is rarely relaxed, and usually together with assumption A3 for the specific purpose of evaluating the effect of tissue capacitance, although in principle dispersion can be considered for purely resistive tissue as well \cite{Gaugain2023Quasi-static}.

\subsection{QSA governing equations}

As a consequence of the four underlying assumptions, the E-fields in tissues under QSA exhibit no spatial propagation attenuation or phase variation, remain linear with applied stimulation amplitude, have no temporal phase delay with the current density, and are independent of stimulation frequency and waveform. This independence of the spatial distribution of the E-field and the stimulation waveform is a fundamental characteristic of the quasistatic approach used in neuromodulation modeling. In this context, the governing differential equation for electrical stimulation is Laplace’s equation for the electric potential. For magnetic stimulation, the Maxwell–Faraday equation determines the primary E-field induced by the magnetic field, which is calculated using Ampère’s law under QSA (i.e., the Biot–Savart law), and the secondary E-field is determined from the primary E-field to satisfy QSA current continuity across tissue boundaries or conductivity gradients.

\subsubsection{Current density in tissue volume conductor\\}

We start by deriving the equation that governs the electric current density $\boldsymbol{J}$ in the tissue under the generalized physical case, with no approximations applied. 
The current density, absent magnetizing currents in a non-magnetic volume conductor, consists of the free current and displacement current (including the polarization current) driven by the time-varying E-field, $\boldsymbol{E}(t)$,

\begin{equation}
       \boldsymbol{J} = \bar{\boldsymbol{\sigma}}(\boldsymbol{r}, \omega, \boldsymbol{E},...)\cdot \boldsymbol{E} + \frac{\partial \left[ \bar{\boldsymbol{\varepsilon}}\left(\boldsymbol{r}, \omega, \boldsymbol{E},...\right)\cdot\boldsymbol{E}\right] }{\partial t}.\label{eq1}
\end{equation}

Here, the conductivity $\bar{\boldsymbol{\sigma}}$ and permittivity $\bar{\boldsymbol{\varepsilon}}$ are tensors and functions of position $\boldsymbol{r}$, frequency $\omega$ (directly proportional to the rate of change of the E-field waveform), and E-field or other potential stimulation-related factors such as temperature, electrophysiological state, or neural activity in response to stimulation.
More generally, the current densities, especially the displacement current, may also not respond to the E-field instantaneously and the multiplications in (\ref{eq1}) need to be replaced with temporal convolution to account for such delays.

The linearity assumption A2 removes the dependency of $\bar{\boldsymbol{\sigma}}$ and $\bar{\boldsymbol{\varepsilon}}$ on the E-field and other factors, and hence their indirect dependency on time, allowing the current density to be given by a spatiotemporal frequency-domain generalized Ohm's law:

\begin{equation}
    \boldsymbol{J} = \bar{\boldsymbol{\sigma}}^{*}(\boldsymbol{r}, \omega)\cdot \boldsymbol{E}, \label{eq2}
\end{equation} 
where $ \bar{\boldsymbol{\sigma}}^{*}(\boldsymbol{r}, \omega) = \bar{\boldsymbol{\sigma}}(\boldsymbol{r}, \omega) + j\omega \bar{\boldsymbol{\varepsilon}}(\boldsymbol{r}, \omega) $ is the complex conductivity tensor. 
As the conductivity $\bar{\boldsymbol{\sigma}}^*$ is inhomogeneous depending on the distribution and structure of the tissue, the spatial dependency on $\boldsymbol{r}$ is omitted but is always implicitly considered in the neuromodulation literature. 
In computational models, the conductivities are typically fixed for each tissue (spatial model compartment), but may vary from compartment to compartment (e.g., lower for bone, higher for cerebrospinal fluid). 
Anisotropy \cite{Shahid2014value,Lee2016Comparison} does not affect the following analysis and, unless it is explicitly considered, scalar conductivity $\sigma$ and permittivity $\varepsilon$ are used for simplicity. 
The frequency dependencies of $\sigma$ and $\varepsilon$ can be obtained from models such as Debye and Cole–Cole equations describing macroscopic tissue measurements \cite{Gabriel1996dielectric} or derived from microscopic composite models \cite{Meffin2014Modelling,Tahayori2014Modelling}. 
Note that in some publications \cite{Bossetti2008Analysis,Caussade2023Towards} the tissue properties have been given in a different but equivalent formulation in terms of complex permittivity $\varepsilon^{*}(\omega) = \varepsilon^{'}(\omega) - j\varepsilon^{''}(\omega)$, where $\varepsilon^{''}(\omega) = \sigma(\omega)/\omega$.

In (\ref{eq2}), the time domain solution of the field vectors $\boldsymbol{E}$ and $\boldsymbol{J}$ have the wave propagation phasor term $e^{j\omega t - \gamma R}$. 
Here, R is the distance from r to the source and the propagation constant $\gamma = (j\omega\mu\sigma - \omega^{2}\mu\varepsilon)^{\frac{1}{2}} = \alpha + j\beta$ consists of the attenuation constant $\alpha$ and the phase constant (wave number) $\beta$ \cite{Plonsey1967Considerations, Branston1991Analysis,Bossetti2008Analysis}, with $\mu$ being the permeability. 
Applying QSA to (\ref{eq2}) given the low frequencies used in neuromodulation and the relatively low conductivities of tissues (see section 2.4 and Fig.~\ref{fig2}), assumption A1 drops wave propagation of $\boldsymbol{E}$ and $\boldsymbol{J}$ ($e^{-\gamma R} \approx 1$ with $|\gamma R| \ll 1$). 
Specifically, given $\omega\varepsilon \ll \sigma$ for neural tissues so that $\alpha \approx \beta \approx (\frac{\omega\mu\sigma}{2})^{\frac{1}{2}}$, both $\alpha R \ll 1$ and $\beta R \ll 1$. 
Equivalently, the electromagnetic skin depth ($\delta = 1/\alpha$) and wavelength ($\lambda = 2\pi/\beta \approx 2\pi\delta$) are much larger than the tissue dimensions ($\delta \gg R$ and $\lambda \gg R$). 
Assumption A1 further indicates that the E-field does not contain contributions from self-induction as further discussed in section 2.7. 
Assumption A3 drops the imaginary part of $\sigma^* (\omega\varepsilon \ll \sigma)$, whereas assumption A4 further removes the dependency of $\sigma$ on $\omega$.

Without internal net macroscopic current sources, the current density within the tissue volume is divergence-free:

\begin{equation}
    \nabla\cdot\boldsymbol{J} = \nabla\cdot(\sigma\boldsymbol{E}) = 0. \label{eq3}
\end{equation}

Current continuity on the boundary of two tissue domains 1 and 2 \cite{Plonsey1977Action,Wang2021Physics} results in $\sigma_1\cdot\boldsymbol{E_{1,\perp}} = \sigma_2\cdot\boldsymbol{E_{2,\perp}}$ for the normal component, or specifically $\boldsymbol{E_{1,\perp}} =0$ if domain 2 is non-conductive, such as air or electrode insulation.

\subsubsection{Electrical stimulation\\}

For electrical stimulation, defining the E-field as the gradient of a scalar potential $\varphi$, $\boldsymbol{E} = -\nabla\varphi$, results in the governing equation for electric potential with the real-valued, fixed, and frequency-independent conductivity, which is commonly used in neuromodulation:
\begin{equation}
    	\nabla\cdot[\sigma(-\nabla\varphi)] = 0.	\label{eq4}
\end{equation}
It may be simplified to Laplace’s equation as
\begin{equation}
    \nabla^2 \varphi = 0 	\label{eq5}
\end{equation}
within each homogenous tissue domain (where $\nabla\sigma = 0$).

The sources in electrical stimulation are typically modeled as boundary conditions applying the current or voltage to the interface surfaces between the tissue and the electrodes, with the latter not necessarily represented explicitly \cite{Minhas2011Cutaneous,Pelot2018Modeling}. 
Depending on the conductivity contrast between the electrode material and tissue, current-controlled stimulation pulses can be modeled as either a Neumann boundary condition specifying an applied current density distribution on the interface (which may be non-uniform), or a floating potential boundary specifying the total applied current for which the solver then calculates the non-uniform current density distribution under a constant surface potential that is unknown (hence floating potential). 
Electrodes with voltage-controlled stimulation pulses are modeled as a Dirichlet boundary condition that specifies a constant potential on the interface surface. 
Models should account for the presence of electrode materials, including any conductive rubber, saline-soaked sponges, or electrolyte gel, when they impact the E-field and current density distribution \cite{Saturnino2015On,Pelot2018Modeling}.

\subsubsection{Magnetic stimulation\\}

In magnetic stimulation, the magnetic field produced by the coil follows the waveform of the coil current in a quasistatic manner according to the Biot–Savart law, with the self-induction of the TMS coil and its influence on the current waveform handled in a prior modeling stage. 
This approximation may account for non-uniform current distribution within thick coil windings but disregards the potential presence of nonlinear and time-dependent magnetic characteristics of the coil materials, such as saturation and hysteresis of the ferrite core. 
The E-field induced by a time-varying magnetic field can be described as
\begin{equation}
    \boldsymbol{E} = -\frac{\partial\boldsymbol{A}}{\partial t} - \nabla\varphi,	\label{eq6}
\end{equation}
where $\boldsymbol{A}$ is the magnetic vector potential that is related to the coil magnetic field $\boldsymbol{B}$ by $\boldsymbol{B} = \nabla\times\boldsymbol{A}$. 
Here, despite the temporal variation being considered very slow (relative to typical radio frequencies \cite{Esselle1992Neural}) under QSA, the primary field $\boldsymbol{E_\mathrm{p}} = -\frac{\partial\boldsymbol{A}}{\partial t}$ generated by the forward mutual induction between the coil and tissue is included because the rate-of-change is high due to the very high coil currents (on the order of kiloamperes). 
In contrast, the induced currents in the tissue (less than 1\,A) are several orders of magnitude smaller than the kA-level coil currents. Therefore, their magnetic field and induced E-field---both self-induction in the tissue and backward mutual induction to the coil---are negligible. The secondary field $\boldsymbol{E_\mathrm{s}} = -\nabla\varphi$ comes from the charge densities on tissue boundaries due to imbalance of the current density of the primary E-field, in a process considered instantaneous under QSA \cite{Branston1991Analysis}. 
Given (\ref{eq3}), the scalar potential is related to the vector potential by \cite{Wang1994three-dimensional,Thielscher2011Impact, Makarov2018Quasi-Static, Gomez2020Conditions}
\begin{equation}
    \nabla\cdot(\sigma\nabla\varphi) = -\nabla\cdot(\sigma\frac{\partial\boldsymbol{A}}{\partial t}) = -\nabla\cdot(\sigma\boldsymbol{E_\mathrm{p}})	\label{eq7}
\end{equation}
and can be obtained numerically via techniques such as the finite- or boundary-element method (FEM or BEM, respectively).

In magnetic stimulation, the presence of the volume conductor comprising tissues does not significantly affect the electrical characteristics of the coils or their magnetic field. 
Therefore, the coil field can be pre-calculated or measured in the form of the magnetic field ($\boldsymbol{B}$ or $\boldsymbol{A}$) or the primary field $\boldsymbol{E}_\mathrm{p}$ and applied as a background field in the numeric solver \cite{Thielscher2011Impact,Makaroff2023Modeling}. 
The field solution is computed for either a specific rate-of-change of the coil current (e.g., 1\,A/\textmu s) or a value to match the coil current waveform characteristics. 
Hence, the electromagnetic induction is handled in a stage preceding the quasistatic modeling of the E-field. 
Alternatively, the coil(s) can be explicitly included as part of the tissue E-field model, and a frequency-domain simulation may be used to calculate the coil magnetic field with the appropriate current amplitude and frequency to reflect the coil current’s rate of change \cite{Wang2018Redesigning}.

\subsection{Frequency spectrum of stimulation pulses and frequency dependence of tissue electric parameters}

\begin{figure*}[p]
	\centering
		\includegraphics[width = \textwidth]{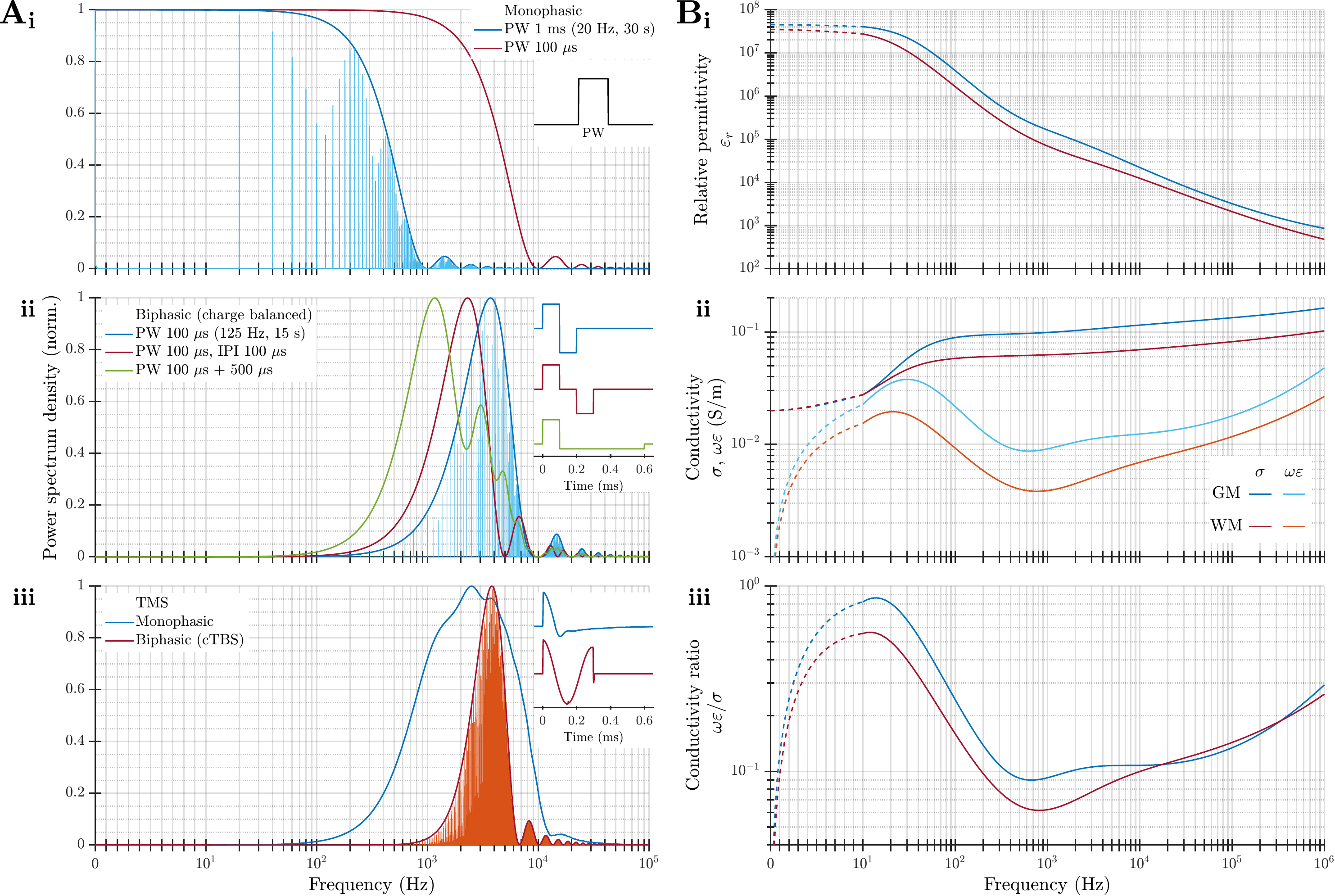}
	  \caption{\textbf{A}.~Normalized power spectrum density between 0 and 100\,kHz of representative neurostimulation pulse waveforms, calculated using a 1~\textmu s sampling rate and a total time interval of 1 s for single pulses. \textbf{i}.~Monophasic pulses with pulse width (PW) of 1~ms (\textit{blue}: single pulse; \textit{light blue}: 30~s pulse train at 20~Hz) and 100~\textmu s (\textit{red}: single pulse). \textbf{ii}.~Charge-balanced biphasic pulses with a first phase of 100~\textmu s duration and with a second phase of equal duration (\textit{blue}: single pulse; \textit{light blue}: 15~s pulse train at 125~Hz), a second phase of equal duration and with 100~\textmu s interphase/intrapulse interval (IPI) (\textit{red}: single pulse), and second phase with 500~\textmu s duration and 20\% amplitude of first phase (\textit{green}: single pulse). \textbf{iii}.~E-field waveforms induced by magnetic stimulation pulses, including monophasic (\textit{blue}: single pulse) and biphasic (\textit{red}: single pulse; \textit{orange}: continuous theta burst stimulation (cTBS), 3 pulses per burst, 50\,Hz intraburst frequency, 5\,Hz interburst frequency, and 40~s duration), recorded from a MagPro X100 device with a Mag-Venture MCF-B70 figure-of-8 coil (MagVenture A/S, Farum, Denmark) \cite{Wang2018Coupling,Aberra2020Simulation}. Electromagnetically-induced pulses are automatically charged-balanced as the amplitude of the coil current returns to zero at the end of a pulse. \textbf{B}.~Spectra of tissue electrical parameters between 0 and 1\,MHz for gray matter (blue) and white matter (red) as modeled by Gabriel \textit{et al.}~\cite{Gabriel1996dielectric}. \textbf{i}.~Relative permittivity. \textbf{ii}.~Resistive (darker colors) and capacitive (lighter colors) conductivities. \textbf{iii}.~Ratio between conductive and capacitive conductivities. Conductivity variations are small over the frequency ranges of charge-balanced pulsed stimulation, enabling the approximation using single conductivity values. Dashed lines indicate extrapolation of the model to very low frequencies ($<10$~Hz), outside the range of experimental data (10~Hz to 20~GHz) to which the model was fitted, where the model has unknown validity. Moreover, the measurements had high uncertainties in the low frequency range (10 to 100\,Hz) due to the experimental setup. All horizontal axes use a mixed scale, with linear scale between 0 and 10\,Hz and logarithmic scale above 10\,Hz.}\label{fig2} 
\end{figure*}

QSA relies on specific electrical parameters of biological tissue, namely the attenuation and phase constants (or their respective inverse, the skin depth and wavelength), the ratio between capacitive and resistive conductances, and how much they vary as a function of frequency. Therefore, it is necessary to evaluate the spectra of these parameters in the frequency ranges relevant to stimulation waveforms used in neuromodulation (Fig.~\ref{fig2}).

tDCS applies dc current via scalp electrodes (disregarding slow ramp-ups and ramp-downs). Usually, tACS is applied using only a single frequency. When the stimulation waveform is a relatively high frequency sinusoid that is amplitude modulated, the frequency spectrum is centered around the carrier frequency---not around the lower frequency of the amplitude modulation. This is also the case for TIS, where the frequency spectrum consists of the relatively higher frequency of the two (or more) individual channels and does not contain low frequency energy \cite{Dmochowski2017Noninvasive,Mirzakhalili2020Biophysics,Wang2023Responses}.

The frequency spectrum of pulsed stimulation depends on the pulse width, shape, and repetition patterns. While monophasic pulses cover a wide frequency range over a few decades starting from dc (Fig.~\ref{fig2}~A\textsubscript{i}), neural stimulation almost always uses charge balanced stimulation pulses (Fig.~\ref{fig2}~A\textsubscript{ii}--A\textsubscript{iii}) that have energy typically distributed within a frequency range of around one decade. The center location of the spectrum is determined by the pulse duration, typically falling between 100\,Hz and 100\,kHz. Increasing the total pulse duration--whether by increasing the phase duration, including an interphase interval, or using an asymmetric pulse with a longer and lower amplitude secondary phase--reduces the center frequency of the spectrum but has smaller effects on its shape. The pattern of pulse stimulation (e.g., continuous versus burst) also does not typically change the overall range or shape of the spectrum as governed by the individual pulses, but rather results in a discretized spectrum with the sampling and envelope determined by the repetition pattern and total duration of the pulse train. The spectra of the stimulation waveform should be considered when selecting the conductivity of tissue(s) within neuromodulation models.

In 1996, Gabriel \textit{et al.} published a series of papers on the electrical conductivity and permittivity of tissue at frequencies from 10\,Hz up to 20\,GHz \cite{Gabriel1996aDielectric, Gabriel1996bDielectric,Gabriel1996dielectric}; this series was subsequently updated \cite{Gabriel2009Electrical}. 
Their data and models, which fit particularly well at high frequencies, are broadly consistent with the dielectric properties measured in other studies and as described by the Cole--Cole equation, and have, to some extent, become a canonical reference for modeling studies explicitly considering the suitability of QSA.

Assumptions A1 and A3 of neuromodulation QSA require tissue conductivity to be both sufficiently low (compared to good conductors like metal) for wave propagation effects to be negligible and sufficiently high (compared to dielectric materials) so that the resistive component of the current dominates compared to the capacitive component. Per the tissue model of Gabriel \textit{et al.}, the attenuation and phase constants for neural tissues are very small, on the order of $10^{-3}$ to $10^{-1}$~Np/m (logarithmic units per meter) and $10^{-3}$ to $10^{-1}$~rad/m, respectively, in the relevant frequency range of 100\,Hz to 100\,kHz \cite{Bossetti2008Analysis}. 
This corresponds to large skin depth and long wavelength on the order of 10\,m even at the highest frequencies used. Thus, wave propagation can be ignored under assumption A1.

According to the data by Gabriel \textit{et al.}, the capacitive current of neural tissues in the relevant frequency range is $\sim 10\%$ compared to the resistive current (Fig.\ref{fig2}~B). 
The total current density consists of the resistive and capacitive currents in parallel and is driven by the same E-field. Because of the 90-degree phase shift between the current components, the total current density is only $\sim 0.5\%$ larger than its resistive component, with a small phase difference of around $6\degree$ between them (and between the current density and E-field). 
Although the modeled capacitive conductance reaches more than half the amplitude of the resistive conductance at frequencies below 100\,Hz, these (extrapolated) values are considered conservative and have high uncertainties in the low frequency range due to the limitation of the measurement setup. 
Further, this low frequency range constitutes a small fraction of the spectrum. Even within the wider range from dc to 1\,MHz, the data by Gabriel \textit{et al.} indicate the average capacitive-to-resistive conductance ratio is $\sim 20\%$, introducing only a $\sim 2\%$ amplitude error and $\sim 11\degree$ phase difference in the current density. 
Such small errors may accumulate for high pulse rate monophasic stimulation, in which the capacitive charge does not have time to dissipate between pulses \cite{Caussade2023Towards}, but such accumulation is mitigated by charge balanced biphasic pulses or low pulse repetition rates. 

More contemporary studies of \textit{in vivo} cortical impedance varied in their conclusions. 
A study that focuses on distortion of endogenous electrophysiological signals (like the initial work by Plonsey and Heppner \cite{Plonsey1967Considerations}) suggests that impedance is largely independent of frequency and can be represented by a purely resistive conductor \cite{Logothetis2007In}.
However, another \textit{in vivo} measurement of tissue permittivity differs substantially from the \textit{ex vivo} data by Gabriel \textit{et al.}; the corresponding models show that the amplitude of capacitive current can reach a significant portion compared to the resistive current for various stimulation waveforms of TMS and DBS \cite{Wagner2014Impact}.
Therefore, based on some application-specific experimental reports, ignoring tissue capacitance under assumption A3 is the weakest approximation \cite{Plonsey1967Considerations,Bossetti2008Analysis}.

The variations of the resistive conductance of neural tissue are generally considered relatively small over the limited frequency range of a decade for charge-balanced pulsed stimulation with a center frequency above 100\,Hz. 
Per the data by Gabriel~\textit{et~al.}, for example, the conductivity of gray and white matter increases from 0.09 to 0.13~S/m and from 0.06 to 0.08~S/m, respectively, over three decades (Fig.~\ref{fig2}~B), corresponding to a change of $\sim 13\%$ per decade on average  \cite{Gabriel1996aDielectric,Bossetti2008Analysis}.
Accordingly, the use of non-dispersive fixed brain tissue conductivity values in most neuromodulation modeling papers to represent the parameters near the center frequency of the pulse is generally considered a valid approximation under assumption A4. 
However, the selection of the fixed frequency and of the corresponding conductivity can have profound effects on the modeled E-field and predicted activation thresholds, due to the sensitivity of the solutions to the conductivity \cite{Bossetti2008Analysis,Caussade2023Towards}.

\subsection{Separability of field solution into product of static spatial distribution and dynamic temporal waveform under QSA}

\begin{figure*}[!p]
	\centering
		\includegraphics[width = 10.6 cm]{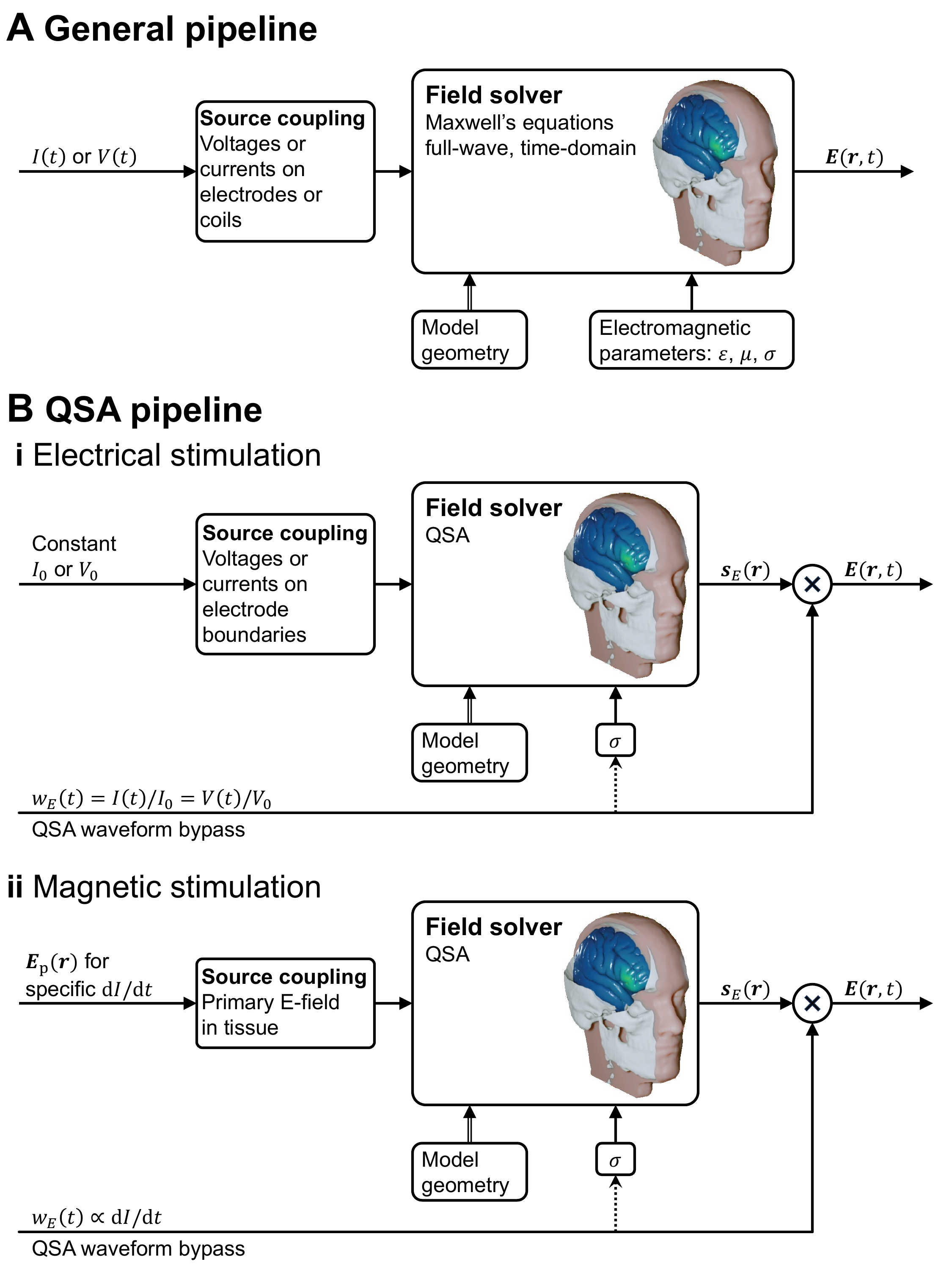}
	  \caption{\textbf{Separability of the electric field spatial distribution and temporal waveform in typical QSA modeling pipelines. A.}~A general field solver, without QSA, couples the calculation of spatial and temporal components. \textbf{B}.~Modeling pipelines based on QSA decouple the step of calculation of the electric field spatial distribution from the waveform. This decoupling occurs at the step of calculating the tissue E-field (see Fig.\ref{fig1} for overall multi-stage modeling pipeline). For electrical stimulation (\textbf{i}), a fixed current $I_0$ or voltage $V_0$ (e.g., corresponding to the waveform peak) is applied to the E-field QSA solver. The QSA solver is thus waveform agnostic and static. The electrical stimulation waveform in effect ``bypasses'' this stage and can be re-coupled to the solved tissue E-field for subsequent modeling steps. The waveform is considered in the QSA E-field calculation only to inform the selection of the fixed tissue conductivity to each tissue volume (dashed arrows). If QSA is applicable to the electrodes as well as the tissues, the waveforms of the electrode current $I(t)$ and voltage $V(t)$ are the same. The process is similar for magnetic stimulation (\textbf{ii}), except that the source is applied in the form of the primary E-field distribution and its magnitude and waveform are determined by the coil current's derivative (rate of change) $\mathrm{d}I/\mathrm{d}t$.}\label{fig3} 
\end{figure*}

One corollary of QSA in neuromodulation is that ``all field components will have the same temporal behavior, i.e., will be in synchrony'' \cite{Plonsey1967Considerations}, and therefore the spatial distribution and temporal waveform of the field are separable \cite{Roth1990model,Ruffini2013Transcranial,Wang2021Physics,Wang2023Responses}.
Specifically, assumptions A1 and A2 result in separability of the spatial and temporal components of the field for any single frequency, whereas assumptions A3 and A4 extend the separability of single-frequency or narrow-band waveforms to arbitrary broadband waveforms. 
The separability leads to decomposing mathematically the magnetic field and E-field into

\begin{numparts}\label{eq8}
\begin{eqnarray}
\boldsymbol{B}(\boldsymbol{r},t) = \boldsymbol{s}_B(\boldsymbol{r})\cdot w_B(t) \; \mathrm{and} \label{eq8a}\\
  \boldsymbol{E}(\boldsymbol{r},t) = \boldsymbol{s}_E(\boldsymbol{r}) \cdot w_E(t), \label{eq8b}
\end{eqnarray}
\end{numparts}
respectively, where $\boldsymbol{s}_B$ and $\boldsymbol{s}_E$ are static spatial vector distributions for coordinates $\boldsymbol{r}$ and are solved for a specific stimulation amplitude, and $w_B$ and $w_E$ are the normalized temporal waveforms (see Fig.~\ref{fig3}). 
The field magnitude and spatial distribution are typically solved with a numeric approach such as FEM or BEM, whereas the waveforms are determined by specifications, circuit simulations, or device output recordings. 
For stimulation via electromagnetic induction, $w_B$ is equal to the coil current waveform and $w_E$ corresponds to its time derivative. 
For stimulation via electrodes, $w_E$ is equal to the current waveform injected through the electrodes and the magnetic fields are not considered. 
If the electrode models are subject to QSA as well, then $w_E$ also matches the inter-electrode voltage waveform. 
Analogously, if the electromagnetic induction coil behaves like an ideal inductor without series resistance, then $w_E$ also matches the coil voltage waveform.

The separability of the spatial and temporal components of the field under QSA has fundamental implications for the implementation of neuromodulation models, and is the key reason why QSA is commonly deployed. 
Specifically, the calculation of the field is time-agnostic since its spatial distribution does not vary over time; only its overall amplitude is scaled multiplicatively by the stimulus waveform. 
Thus, field solvers need only to compute $\boldsymbol{s}_B(\boldsymbol{r})$ or $\boldsymbol{s}_E(\boldsymbol{r})$ as a static field map. 
The waveform is factored in the static field solution only insofar as it informs the selection of fixed tissue conductivity, for example to account for the frequency content of the stimulus. 
Other than this, the waveform ``bypasses'' the field solver, only to be combined with the spatial distribution in later modeling stages (Fig.~\ref{fig3}), for example when calculating neuronal stimulation in multistage modeling (Fig.~\ref{fig1}).

Applications using multiple stimulation channels with different temporal waveforms for each channel may require special considerations, since the spatial distribution of the total E-field is no longer static despite being a (time-dependent) linear sum of the individual static fields. 
Examples of such methods include amplitude or pulse-width modulated TIS \cite{Grossman2017Noninvasive,Sorkhabi2020Temporally,Howell2021Feasibility,Missey2021Orientation,Acerbo2022Focal,Wang2023Responses}, multi-phase/traveling-wave multi-channel tACS using the same frequency but different phase shifts \cite{SATURNINO201768,Alekseichuk2019Electric,Lee2023Predicting,Wang2023Responses}, ISPS applying asynchronous or interleaved pulses across multiple channels to achieve stimulation where the high-amplitude regions of individual E-field distributions intersect \cite{Vöröslakos2018Direct,Howell2021Feasibility}, and rotational-field TMS \cite{Rotem2014Solving,Roth2020Rotational,Roth2023Revisiting} using two TMS coils with a quarter-cycle delay between the two cosinusoidal pulses. 
QSA can be applied to solve the E-fields due to each source, and then the solutions can be linearly combined to obtain the total field

\begin{equation}
    \boldsymbol{E}(\boldsymbol{r},t) = \sum_l \boldsymbol{s}_{E,l}(\boldsymbol{r}) \cdot w_{E,l}(t),
\end{equation}
where $l$ enumerates the channels. 
The total E-field of such stimulation methods exhibits additional features of potential interest, including global traveling waves \cite{Alekseichuk2019Electric,Lee2023Experimental} and local rotation \cite{Rotem2014Solving,Wang2023Responses}, requiring additional care for subsequent field analysis and neural simulation, such as phasor analysis \cite{Lee2023Predicting} and realistic neuronal morphology to capture its interaction with the non-quasistatic total E-field \cite{Wang2023Responses}.

\subsection{Adaptation of QSA for relaxing individual assumptions}

The application of governing equations (\ref{eq4}) and (\ref{eq7}) in the specified form---with $\sigma$ a real-valued constant, which is frequency-independent and linear--inherently satisfies all four underlying assumptions of QSA in neuromodulation.
In this section, we consider how, in the context of the ubiquitous use of QSA in neuromodulation, there are modeling pipelines that consider frequency-, time-, or field-dependent tissue properties, while still satisfying QSA for each E-field calculation (Fig.~\ref{fig1}).
These scenarios reflect relaxation of one or more of assumptions A1--A4 at the level of the modeling pipeline (or more generally study) scale, but all QSA assumptions remain at each (iteration of) tissue E-field modeling. 
As is the case elsewhere in this document, our scope here is not to comment on the application-specific preference between these approaches, but explain how to distinguish them.

At a most general level, a given study could encompass multiple simulation scenarios, each involving distinct $\sigma$ parameters (such as exploring the impact of tissue conductivity or comparing subject-specific conductivity).
Nevertheless, as long as each scenario maintains consistent fixed values of $\sigma$, then QSA is upheld.

For electrical stimulation, areas of particular concern regarding frequency-, time-, and intensity-dependent impedances relate to the capacitance of the tissue and electrode \cite{Butson2005Tissue,Grant2010Effect,Gaugain2023Quasi-static} and nonlinear behavior of the electrode interface \cite{Cantrell2008Incorporation} and the skin \cite{Panescu1994nonlinear,Unal2021Adaptive}. 
It is notable that for the case of skin, none of the assumptions A2, A3, or A4 holds (see sections 3.4 and 3.5).
As the simplest adaptation, most transdermal/transcranial electrical stimulation models adopt QSA by (supposedly) selecting skin conductivity that best accounts for the skin’s complex behavior; the validity of QSA is further supported by the fact that most tES applications employ current-controlled stimulators, which reduces the sensitivity of the delivered E-field to the dynamic aspects of the skin conductivity.
For magnetic stimulation (see section 3.6), deviation from QSA mostly comes from the stimulation coil, such as the frequency-dependent skin effect in thick conductors in coil windings \cite{Gomez2020Conditions} or the nonlinearity of ferromagnetic core materials \cite{Makaroff2023Modeling}.

\begin{figure*}[!p]
	\centering
		\includegraphics[width = 12.4 cm]{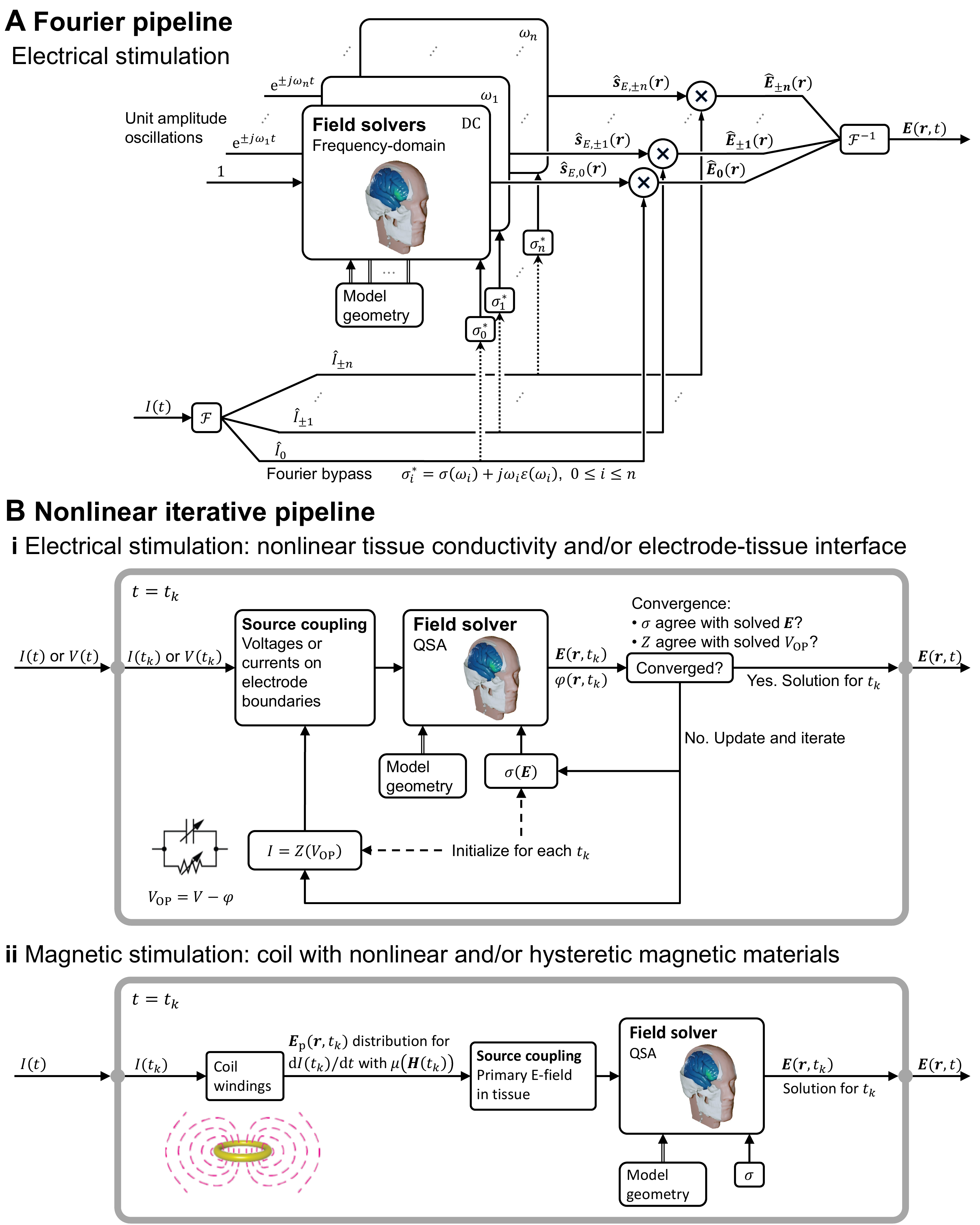}
	  \caption{\textbf{Adaptations of QSA framework. A.}~A Fourier pipeline solves the E-field in response to an arbitrary stimulation waveform in dispersive tissue by decomposing the waveform into its frequency components, with the number of the components $n$ sufficient to capture the waveform spectrum. Each component is solved individually under QSA using frequency-specific tissue parameters. The pipeline can also be used for capacitive tissue and/or for obtaining full-wave solutions, for which individual field solutions are no longer under QSA. Symbols with hats indicate frequency domain variables. \textbf{B.}~Iterative pipelines solve nonlinear systems. \textbf{i.}~For electrical stimulation, tissue conductivities can be a nonlinear function of the E-field, and electrode current and voltage can be related by a non-linear electrode-tissue interface via the overpotential $V_{\mathrm{OP}}$ of the interface. \textbf{ii.}~For magnetic stimulation using coils with nonlinear magnetic materials, the permeability is a function of the magnetic field intensity $\boldsymbol{H}$, and the primary E-field depends nonlinearly on the coil current. The gray boxes indicate that the computation is performed for a discrete time step. In some cases, information needs to be transferred from one step to the next, e.g., for capacitive electrode interface and hysteretic magnetic core.}\label{fig4} 
\end{figure*}

If the key assumption of linearity (A2) is held, the relaxation of the other assumptions can be handled using the frequency domain.
When tissue dispersion is considered (relaxing assumption A4) so that $\sigma$ is subject to variation of the frequency content of the waveform---be it through tabulation or functional definition---the field solution becomes a function of frequency.
The linearity assumption allows QSA to “apply to a linear combination of harmonic frequencies and hence, to a periodic or to an aperiodic source through the use of a Fourier series or integral, respectively” \cite{Plonsey1967Considerations}.
For either periodic or aperiodic waveforms, this is typically approached by decomposing the problem into a finite series of harmonic (frequency-specific) QSA solutions using the discrete Fourier/Laplace transform.
Namely, the waveform is decomposed into a set of sinusoids (single frequencies), the E-fields are calculated for each sinusoid (which, with A4 relaxed, involve frequency-specific tissue properties), and the set of resultant E-fields are re-combined by linear superposition. 
This reflects an adaptation (Fig.~\ref{fig4}~A) since QSA is still applied for each E-field calculation.

This same Fourier framework can still be used for non-QSA cases where 1) tissue capacitance is included (relaxing assumption A3) and the phase delay between the current density and E-field introduces another frequency dependency regardless of tissue capacitance being dispersive \cite{Grant2010Effect,Gaugain2023Quasi-static} or not \cite{Butson2005Tissue}, and 2) wave propagation is included (relaxing assumption A1) and the full-wave solution to the Helmholtz equation is obtained, as typically performed for validation of QSA \cite{Bossetti2008Analysis,Caussade2023Towards,Gaugain2023Quasi-static}. 
When tissue (and electrode) capacitance is included, the transfer function between the E-field and stimulation sources could have multiple poles and zeros due to different time constants in the various tissue compartments \cite{Bora2020Estimation}, resulting in complex temporal behavior of the E-field waveform.

The most significant challenge for QSA in neuromodulation is when the nonlinearity assumption (A2) is relaxed.
Assumptions A1, A2, and A4 can still be applied to the inhomogeneous electromagnetic wave equation and the current density constitutive equation in the time domain to neglect wave propagation ($|\mu\varepsilon \frac{\partial^2\boldsymbol{E}}{\partial t^2} + \mu\sigma \frac{\partial\boldsymbol{E}}{\partial t}| \ll |\nabla^2\boldsymbol{E}|$) and capacitive tissue current ($|\frac{\partial(\varepsilon\boldsymbol{E})}{\partial t}| \ll |\sigma\boldsymbol{E}|$), respectively \cite{Branston1991Analysis,Bossetti2008Analysis}. 
With assumption A1 alone, however, the solution no longer has separability of time and space, as the spatial distribution is dependent on the instantaneous amplitude of the stimulation waveform, i.e., the relative distribution varies with time.
The system needs to be analyzed for each time “snapshot” of the waveform.

This has been considered in detail for ECT, which applies high current across the scalp, a tissue that demonstrates nonlinear impedance \cite{Unal2021Adaptive,Unal2023Quasi-static}. 
For example, locations near the scalp electrodes experience the nonlinear effect earlier as the high local E-field changes the skin properties.
Nevertheless, an iterative approach that preserves QSA at the stage of each E-field calculation can be adopted (Fig.~\ref{fig4}~B\textsubscript{i}) \cite{Panescu1994nonlinear,Unal2021Adaptive,Unal2023Quasi-static}.
Conductivity values are initially assigned agnostically, and once the field is solved (under QSA), it is used to update the conductivity iteratively, until the E-field amplitude in the tissue used for selecting the conductivity matches the solved E-field. 
This process assumes the tissue conductivity instantly adjusted to the local tissue E-field and can be done in separate simulations for any given stimulation dose or waveform amplitude. 
A nonlinear electrode–tissue interface, whether modeled with lumped or distributed parameters, can be similarly processed \cite{Cantrell2008Incorporation}. 
The impact of explicitly modeling electrode interfaces will depend on the application, how the interface is represented, and the model output.

In TMS, when nonlinear (saturable and/or hysteretic) magnetic core materials are present in the coil, the magnetic field (either B or A) and hence the primary E-field are similarly not separable into individual spatial and temporal components. 
The primary E-field distribution needs to be calculated at each waveform snapshot \cite{Makaroff2023Modeling} allowing the corresponding secondary E-field to be solved under QSA (Fig.~\ref{fig4}~B\textsubscript{ii}). 
Although the change in the relative spatial distribution of the total E-field in the brain may be limited, due to its distance to the nonlinear core, the E-field waveform will experience unique distortion resulting from the simultaneous increase in the coil current due to the drop in inductance combined with the reduction of the magnetic field output as a result of core saturation: the E-field’s peak amplitudes, corresponding to the largest rate of change of the magnetic field, typically occur when the magnetic field has small amplitudes, such as at the pulse start or when the magnetic field flips polarity; in contrast, the peak amplitudes of the magnetic field, at which the core material saturates, occur at different times during the pulse and typically correspond to a small E-field \cite{Epstein2002Iron-Core,Makaroff2023Modeling,Nguyen2023High}.

Effects of tissue heating and other slow (relative to the pulse and pulse train) and nonlinear processes can also be analyzed using a QSA pipeline by expanding the time scale into decoupled fast and slow scales that are iteratively solved \cite{Krassowska1994Response, Cranford2012Asymptotic}. 
The fast (sub-second to second) scale uses QSA to calculate field distribution for individual pulses, which are then used to determine effects on the slow (minutes to hours) scale such as heating \cite{Datta2009Bio-heat,Khadka2020Bio-Heat,Zannou2021Tissue,Zannou2023Bioheat}.
These effects in turn update the system in the QSA framework, e.g., with temperature-dependent tissue properties \cite{Gholami-Boroujeny2015Theoretical,Wang2024Optimizing}.

Such an iterative QSA framework can also be used for microscopic field modeling that includes membrane capacitance and nonlinear ion channels in addition to membrane resistance. 
In this context, QSA is considered in the intra- and extracellular domains and the four assumptions remain valid \cite{Lindsay2004From} when evaluated on the smaller cellular dimension and with electromagnetic parameters for the intra- and extracellular fluids, even considering a shorter time scale (higher frequency) necessary to resolve the fast dynamics of microscopic transcellular membrane charging with time constants on the order of 10–100\,ns (10–100\,MHz) \cite{Cartee1992transient,Krassowska1994Response,Ying2007Hybrid,Agudelo-Toro2013Computationally,Wang2018Modified}. 
The dielectric contribution of the cell membranes, which are the main source of the tissue’s macroscopic capacitance in the relevant frequency range \cite{Meffin2012Modeling, Meffin2014Modelling,Tahayori2012Modeling,Tahayori2014Modelling}, and the nonlinear ion channels are confined to the boundaries between the intra- and extracellular domains as a boundary condition of discontinuous voltage with capacitive and nonlinear transmembrane current.

\subsection{QSA from the perspective of computational electromagnetics}

It is important to note that the interpretation of quasistatic assumptions so far is specific to the domain of neuromodulation. 
QSA as implicitly indicated in neuromodulation modeling studies is distinct and typically stricter than the use of this term, and analogous ones, in the broader field of electromagnetics. 
For example, a “quasistatic” numerical solver in a software package \cite{Cantrell2008Incorporation} typically reflects the use of this term in electromagnetics, as would descriptions such as “electrostatic double layer capacitance”. 
Here, the differences with general electromagnetics are elaborated, where at least two QSAs of Maxwell’s equations are used depending on the electromagnetic field source types and relative importance of coupling between the electric and magnetic fields. 

Maxwell's equations decouple into electrostatic and magnetostatic formulations when all time derivatives are unimportant and can be neglected. 
In contrast, when one of the time derivatives becomes significant, electro-quasistatic (EQS) and magneto-quasistatic (also denoted as magneto-quasistationary; MQS) systems arise \cite{Haus1989Electromagnetic, Steinmetz2009Domains}. 
It should be emphasized that, while both EQS and MQS approximations neglect electromagnetic wave propagation as described under assumption A1, they respectively neglect either the electromagnetic induction (EQS) or displacement current (MQS), but not both together.

As an example of EQS application, the E-field in the brain during tES is not static but evolves over time. 
Consequently, it generates a solenoidal (divergence-free) magnetic field, which, in turn, produces a new E-field that opposes the initial one. 
However, due to the small ratio of the head size to the sub-MHz range wavelengths of the stimulation waveforms, even when considering the highest harmonics, the energy stored in the magnetic field is inconsequential \cite{Gaugain2023Quasi-static}. 
As a result, this E-field correction term by the magnetic field has an amplitude proportional to $|\gamma R|^2$ when compared to the E-field generated by the electrode charges, and is thus negligible as long as propagation is neglected \cite{Plonsey1967Considerations}. 
This allows for a simplified modeling approach that focuses on the time-independent component of the E-field from the "zeroth-order" laws, whereas the first-order magnetic field, the less important “left-over” quantity in EQS \cite{Haus1989Electromagnetic}, is of no interest to neuromodulation and simply ignored.

On the other hand, TMS operates on the principle that a changing magnetic field will induce an E-field, as described by the Maxwell–Faraday equation. 
The electric current due to the induced E-field, in turn, generates an opposing magnetic field. 
Within the MQS approximation, this correction term to the magnetic field by the tissue current is neglected, as it is similarly proportional to $|\gamma R|^2$ when compared to the magnetic field generated by the coil currents \cite{Polk1990Electric, Wang1994three-dimensional}. 
This allows the magnetic field and hence the divergence-free component of the total E-field (i.e., the primary E-field) to be calculated directly from the TMS coil current. 
However, the secondary E-field resulting from the boundary charge density distribution---the MQS “left-over” quantities \cite{Haus1989Electromagnetic}---still needs to be calculated to arrive at the total E-field \cite{Plonsey1967Considerations,Thielscher2011Impact, Romero2019Neural}, which determines neuromodulation dose and neuronal responses \cite{Peterchev2012Fundamentals}.

The electrostatic approximation in electromagnetics also makes the assumptions A1, A3, and A4 of QSA in neuromodulation. 
While assumption A2 (linear tissue properties) is not directly related to the static approximation, it can be applied to the static approximation depending on the specific problem under study, pertaining to the linearity of its governing equation and boundary conditions. 
Therefore, the term “(electro)static approximation” has been used interchangeably with QSA in electrical stimulation \cite{Butson2005Tissue,Foutz2010Evaluation,Gaugain2023Quasi-static}. 
However, such usage is rare and may lead to confusion.
In electromagnetics (and physics more broadly), the term "static" implies that all variations in time (or time derivatives) of the initial and boundary conditions are considered negligible. 
Therefore, only direct current stimulation neglecting any ramps or tissue adaptation would strictly qualify as static from an electromagnetics perspective, whereas a system described by Laplace’s equation alone does not. 
Furthermore, factors such as nonlinear tissue properties (referred to as adaptive in the literature, e.g., a slow change in skin conductivity during tDCS) can be accurately represented under the static pipeline, even though the solution evolves on the longer time scale. 
Therefore, referring to neuromodulation models as (electro)static or using electrostatic approximation is a more conscientious application of the four assumptions, and the term "quasistatic" is more commonly used in the broader neuromodulation literature, except in rare cases when researchers explicitly study the differences between both approximations \cite{Gaugain2023Quasi-static}.

\section{Historical adoption and analysis of QSA in neuromodulation}

\subsection{Scope of historical review}

We briefly summarize the historical adoption of QSA, starting with early and general developments and then in specific applications by neuromodulation domain. 
As is the case throughout this review, our focus is on clarifying how QSA is used and not judging its appropriateness, although examples of validation efforts are noted. 
The conclusions of any examples evaluating the validity of the QSA depend on:

\begin{enumerate}
    \item The specific conditions, scenarios, or applications: Findings from studies on one modality (e.g., microelectrode stimulation of an axon) may not apply to another (e.g., TMS).
    \item The specific assumptions tested: Some assumptions may not be explicit and even the meaning of “QSA” may vary between reports. When QSA simulations are compared with those under relaxed assumptions, any comparisons are themselves entirely dependent on the assumed tissue properties (e.g., the importance of dispersion depends on the assumption of the magnitude of dispersion).
    \item The modeling methods: The E-field errors introduced by QSA cannot be understood without considering how each model is used. Any change in modeling methods will impact E-field prediction to some extent, so that the meaning of errors depends on how these translate to differences in neuromodulation practice (e.g., which dose is selected) or how mechanisms are interpreted (e.g., predicted neuronal response).
\end{enumerate}

\subsection{Early and general development of QSA}

By comparing frequency-dependent resistive versus capacitive tissue currents, experimental work by the 1950s \cite{Cole1941Longitudinal,Schwan1955Electrical,Schwan1957Capacitive, Schwan1957Conductivity} started to describe conductivity of biological tissues or media as “purely resistive” for applications with frequencies up to 1\,kHz. 
Plonsey and Heppner \cite{Plonsey1967Considerations} provided a complete analytical framework. 
Despite only mentioning non-neural tissue and only considering bioelectric sources, their work became canonical in neuromodulation research in support of QSA \cite{Bossetti2008Analysis}. 
Plonsey and Heppner start with linear tissue properties (A2) and consider signals of (endogenous) physiological origin having frequencies below 1\,kHz to conclude that assumptions A1 and A3 (no electromagnetic wave or induction and resistive tissue) are justified.
Assumption A4 (frequency-independent conductivity) was not considered. 
Plonsey and Heppner note “[i]n the quasi-static approximation, a purely resistive medium is required”--and the association of “purely resistive” tissue with QSA repeats. 
In comparison, we define QSA as conventionally used in neuromodulation to imply all four QSA assumptions. 
Further, Plonsey and Heppner implied EQS and MQS (see section 2.7), but this is not conventional terminology in neuromodulation modeling. 
Plonsey, in subsequent studies \cite{Plonsey1977Action, Plonsey1982nature,Plonsey1984Quantitative, Plonsey1988Bioelectric}, continued to describe biological tissue/medium and their associated fields using the terms “resistive” and “quasistatic”, respectively. 
Since 1996, experimental measurements by Gabriel~\textit{et~al.}~\cite{Gabriel1996bDielectric}(see section 2.4) provided indirect support for QSA, by informing the values used for purely resistive tissue.

Historically, the use of QSA was driven in part by the reliance on analytical or limited numerical solution methods, coupled with simplified geometries such as point electrodes or planar or spherical tissue structures \cite{Rush1968Current,Tranchina1986model,Rattay1988Modeling, Warman1992Modeling}. 
The work by Rush and Driscoll \cite{Rush1968Current} is generally noteworthy for its early attempt to model and experimentally verify current flow models of transcranial electrical stimulation and serves as an example of the implicit adoption of QSA in neuromodulation.
With the introduction of more advanced numerical techniques like FEM and increased computational resources, more intricate electrode and tissue geometries are implemented \cite{McIntyre2004Cellular, Datta2009Gyri-precise, Zhang2010Modeling}. While a small minority of studies have applied non-QSA approaches, QSA remains almost universal at the stage of calculating tissue E-fields.

Significant effort went into developing models of neuronal membrane polarization, action potential generation, and network stimulation (i.e., starting from predicted E-field through prediction of physiological outcomes), including analysis of the importance of neuronal morphology and biophysics \cite{Tranchina1986model, Rattay1999basic}. 
The four QSA assumptions are in the implicit background of these studies, when either calculating (e.g., FEM or analytical) or assuming (e.g., quasi-uniform) an imposed E-field. 
In some cases, these reports cite Plonsey and note “purely resistive medium” \cite{Nicholson1973Theoretical, McNeal1976Analysis, Warman1992Modeling}. 
Thus, canonical studies establishing neuronal modeling methods implicitly propagated QSA (in a general sense) in neuromodulation.

Thus, spanning decades of development, models of cell polarization by electrical stimulation (implicitly) adopt the four quasistatic assumptions; this work spans spheroidal cells \cite{Klee1976Stimulation, Kotnik2000Analytical, Lee2005Polarization}, (semi) infinite axons \cite{McNeal1976Analysis,Coburn1985Theoretical2,Rubinstein1993Axon}, and neurons \cite{McIntyre1999Excitation,Joucla2009“Mirror”,Arlotti2012Axon,Aberra2020Simulation}. 
Since the 1970’s, all four quasistatic assumptions seem established (implicitly) across neuromodulation modeling, often by stating “purely resistive” tissue \cite{Nicholson1973Theoretical}. 
For example, in his reviews of neuromodulation of axons, Rattay considers only the “quasistatic case (where biological tissue impedances are purely resistive)” \cite{Rattay1988Modeling}. 
In developing the method whereby intracellular current injection approximates excitation by applied fields, a “purely resistive” tissue is assumed \cite{Warman1992Modeling}.
As more sophisticated models were developed, it became superfluous (forgone assumption) to even mention “purely resistive”---as a surrogate, authors may simply refer to solving Laplace’s equation and provide only conductivity values for the tissue properties \cite{Meffin2012Modeling}.

\subsection{Spinal cord stimulation, DBS and point source (microelectrode) stimulation}

Computational models of spinal cord stimulation (SCS) date from work by Coburn \cite{Coburn1980Electrical}, and include early implementations of FEM models. 
While the term quasistatic is not mentioned in these early works, they assume “steady state conditions” along with the underlying assumption that tissue is “purely resistive” \cite{Coburn1980Electrical}. 
Through the 1980’s and 1990’s, pioneering SCS modeling by Colburn and colleagues, Holsheimer and colleagues, and others, adopted QSA implicitly or by general description of tissue properties (e.g., “no reactive components, ... steady-state, volume conductor field problem”) \cite{Coburn1985Theoretical1}. 
In 1965, Ranck and BeMent measured the impedance of spinal cord dorsal columns, noting clear anisotropy and some frequency dependence (from 5\,Hz to 50\,kHz) \cite{Ranck1965specific}.

Ongoing efforts to model SCS ubiquitously rely on the four quasistatic assumptions; this is generally implicit, and many studies explicitly note the use of Laplace's equation and (purely) resistive tissue properties \cite{Howell2014Evaluation,Khadka2020Realistic}.
This includes models of higher (kHz) frequency stimulation \cite{Lempka2015Computational, Lempka2020Patient-Specific}. 
Temperature increases by high-duty cycle SCS are modeled \cite{Zannou2021Tissue,Zannou2023Bioheat} but not coupled to changes in tissue conductivity, which would challenge assumption A4.

The validity of QSA has been analyzed to a limited extent, relying on the available tissue data (see section 2.4).
Bossetti~\textit{et~al.}~\cite{Bossetti2008Analysis} evaluated QSA for predicting potentials generated by short-duration stimulation pulses (with high-frequency content) for the case of a nerve fiber near a point-source electrode; for the quasistatic case, Bossetti~\textit{et~al.} explicitly noted assumptions A1 and A3 and implicitly included assumptions A2 and A4. 
However, for a non-sinusoidal (pulsed) signal composed of multiple frequencies (harmonics), tissue capacitance may affect the induced E-field \cite{Butson2005Tissue,Bossetti2008Analysis}.

\subsection{tDCS, tACS, CES}

Relevant to all forms of transcranial electrical stimulation, decades of detailed analysis of skin impedance have shown that none of assumptions A2, A3, or A4 holds. 
Rather, skin properties are complex, layer specific, can in cases be described by the Cole–Cole model, and change nonlinearly with intensity \cite{Yamamoto1976Electrical,Bora2020Estimation}. 
This has implications for modeling of transcutaneous electrical stimulation including tES, but is disregarded under (implicit) quasistatic assumptions. 
Nonetheless, these dependencies are evident in tES technologies where impedance varies with applied current \cite{Hahn2013Methods}. 
Only a few FEM approaches consider the dependence of skin conductivity of local E-field \cite{Panescu1994nonlinear, Unal2021Adaptive}.

Models of tDCS adopted the four QSA assumptions. 
Such efforts started with concentric sphere models \cite{Miranda2006Modeling, Dmochowski2012point} and were extended to smooth anatomy rendered from magnetic resonance imaging (MRI) using computer aided design \cite{Wagner2007Transcranial} and contemporary MRI-derived “gyri-precise” anatomy \cite{Datta2009Gyri-precise}.
Under direct current, assumptions A3 and A4 seem evident since the stimulation is not changing with time \cite{Paulus2013Ohm’s}. 
In tDCS studies, QSA may not even be mentioned but is implied by the reduced governing equations (see section 2.3.2). 
Nonetheless, open questions on modeling methods remain, such as which tissue conductivities are appropriate (given that measurements of conductivities rarely use dc test currents) and how they may change with current or time as suggested by the reduction in overall resistance \cite{Hahn2013Methods}.
There have been some validation studies of predicted tDCS current flow \cite{Datta2013Validation,Antal2014Imaging,Esmaeilpour2017Proceedings}, although more extensive evidence may come from validation of tACS, which is considered to be computationally analogous.

Models of tACS were adapted from tDCS pipelines under (implicit) QSA, typically assuming the same tissue conductivity parameters \cite{Neuling2012Finite-Element}. 
Measurements of tACS generated potentials using intracranial electrodes show small but non-negligible changes in tissue E-fields over frequencies of $\sim$ 0.1--150\,Hz \cite{Opitz2016Spatiotemporal, Huang2017Measurements}. 
This can be interpreted as suggesting that assumptions A3 and A4 are reasonable, but indeed approximations.

Reviews on computational modeling of tES confirm ubiquitous use of the quasistatic assumption, but only implicitly or incompletely indicate the four underlying assumptions \cite{Bikson2012Computational}. 
For example, Bai~\textit{et~al.}~\cite{Bai2013Review} note assumptions A1 and A3, whereas Ruffini~\textit{et~al}.~ \cite{Ruffini2013Transcranial} further note the separability of spatial and temporal components of the E-field.

A recent study directly accessed the QSA errors for tACS, with specific clarification of each assumption \cite {Gaugain2023Quasi-static}.
QSA-induced E-field errors were analyzed as a function of frequency in the 10\,Hz to 100\,MHz range, necessitating an implicit relaxation of assumption A4.
Subsequently, assumption A3 was relaxed to examine the impact of tissue capacitance on the error. 
The reference solution employed a general case of the full set of Maxwell’s equations (i.e., assumption A1 was also omitted, and wave effects were considered).
When comparing the results involving tissue capacitance (without assumption A3) to this general solution, the limitation of assumption A1 was identified in the\,MHz range. 
However, assumption A3 was observed to introduce up to a 20\% error in the brain at 10\,Hz.

Modeling pipelines (from tissue segmentation and properties to QSA E-field simulation) were developed for CES and TIS as well \cite{Grossman2017Noninvasive,Rampersad2019Prospects, Esmaeilpour2021Temporal, Gomez-Tames2021Multiscale,Wang2023Responses}.
As these approaches may use higher (carrier) frequencies (e.g., $> 1$\,kHz), the selection and appropriateness of tissue conductivity values should be considered (see section 2.4).

\subsection{Suprathreshold TES and ECT}

Suprathreshold TES and ECT are unique among electrical stimulation techniques in being both high current ($\sim 500\times$ that in tDCS) and applied across the skin.
Modern TES and ECT modeling derives much from modeling tDCS in regard to development of tissue masks, and historically follows QSA \cite{Weaver1976Current,Suihko1998Modeling,Nadeem2003Computation,Deng2011Electric,Edwards2013Physiological,Lee2016Comparison,ahmadbakir_finiteelement_2019,Deng2019Electric}. 
QSA assumption A2 has been questioned for ECT (and, by extension, for suprathreshold TES), as the high currents ($\sim$ 1\,A) produce a decrease in tissue resistivity, especially in the skin, which is indeed the source of the higher static impedance (\textmu A test current) compared to dynamic impedance (800–900\,mA stimulation current) \cite{Unal2021Adaptive}.
The relevance of assumptions A3 and A4 have also been explicitly considered for ECT, leading to a recommendation for a quasistatic pipeline (i.e., individual modeling steps based on QSA) \cite{Unal2023Quasi-static}. 
These recent developments support discussion over the applicability of QSA for ECT as well as additional assumptions such as anisotropy \cite{Sartorius2022Electric,Deng2023On}.

A few early studies undertook measurements of the E-field generated by ECT in human cadavers \cite{Smitt1944On,Lorimer1949Path} and in an electrolytic tank \cite{Rush1968Current}. 
Such efforts are limited by expected tissue property changes following death, including the absence of tissue conductivity changes with high currents. 
Further, these cadaver tissue measurements relied on sinusoidal current as was used historically, with associated frequency characteristics distinct from modern pulsed stimulation. 
For these reasons, these reports did not validate the key issues debated for modern ECT modeling.

\subsection{Magnetic stimulation }

While not directly using the term, the TMS literature in the 1980s implied QSA by analyzing the skin depth and/or ratio between induced tissue current versus coil current \cite{Parkinson1985Electromagnetic,Ueno1987Localized, Tofts1990distribution, Davey1991Prediction, Grandori1991Magnetic}. 
The earliest use of the term “quasistatic” in the context of TMS appears to be by Roth and Basser~\textit{et~al.} \cite{Roth1990model,Roth1990theoretical}, citing Plonsey and Heppner \cite{Plonsey1967Considerations}. 
Both early examples \cite{Saypol1991theoretical, Esselle1992Neural, Heller1992Brain, Murro1992model} and later publications typically cited either Plonsey and Heppner \cite{Plonsey1967Considerations}, Roth and Basser \cite{Roth1990model}, or Grandori and Ravazzani \cite{Grandori1991Magnetic} for QSA. 
Wang and Eisenberg \cite{Wang1994three-dimensional} summarized the assumptions A1, A3, and A4 for QSA (with assumption A2 implicitly included) in the context of developing FEM methods for TMS, and their work is often referenced in subsequent modeling studies and frameworks \cite{delucia_diffusiontensor_2007,Opitz2011How,Thielscher2011Impact,Windhoff2013Electric}. Validation of QSA performed specifically for TMS using frequency-dependent tissue-specific permittivity parameters demonstrated that tissue capacitance is negligible for relative permittivity not exceeding certain limits \cite{Wagner2004Three-dimensional}.

\section{Conclusions and recommendations for communication of QSA in neuromodulation}

Although the use of QSA in neuromodulation modeling is ubiquitous, it is often implicit, and rarely are the specific assumptions articulated. 
Here we make recommendations to improve the rigor and reproducibility of neuromodulation modeling, in the context of QSA. 
The description of a modeling study should be sufficient to support its reproduction \cite{McDougal2016Reproducibility,Bikson2018Rigor,Miłkowski2018Replicability,mulugeta_credibilityreplicability_2018,Janssen2020On}, including by individuals who may not otherwise understand unstated contemporary norms. 
This can be accomplished by detailed reporting of modeling procedures, parameters, boundary conditions, and other model and solver settings, and/or references to prior publications. To this end, this document explicitly defines QSA in neuromodulation.

\begin{enumerate}
    \item When a modeling study applies QSA in a manner consistent with the four assumptions (and associated pipeline) explained here, this should be stated explicitly. Although QSA is the norm in neuromodulation modeling, not making its application explicit and precise can create ambiguity. As clarified here, QSA is applied specifically at the tissue E-field modeling stage.
    \item When a modeling study applies QSA in a manner distinct from the four assumptions (and associated pipeline) explained here, or when non-QSA methods are used, such steps should be described explicitly and a rationale should be provided. However, as described here, there are adaptations (e.g., Fourier analysis or tissue property adaptation outside the tissue E-field modeling steps) that remain within the QSA framework (see section 2.6).
    \item QSA, as explained here, simplifies the tissue electrical parameters to be linear, resistive, and fixed across frequency. Notwithstanding the common practice of “carrying over” parameters from prior modeling studies, it is incumbent in every modeling study to recognize that the results (e.g., neuronal activation thresholds) are strongly dependent on the selection of parameters. As such, under QSA, the selection of a single fixed conductivity for each tissue should be rationalized. This requires selection of a representative frequency---since tissue conductivity is dependent on frequency---based on the frequency spectrum of the stimulation signal (see section 2.4 and Fig.~\ref{fig2}). Whether individual anatomy and/or anisotropy is considered does not impact QSA, although how such tissue electrical properties are parametrized must be detailed.
    \item Recognizing that all models depend on approximations based on evolving knowledge, the question is not whether QSA is “true”, but rather when it is appropriate to the problem and its solution. The suitability of QSA will depend on a) the neuromodulation dose including stimulus position, intensity, and frequency content (i.e., type of stimulation, see section 3) and b) the question being asked, including steps before and/or after the QSA E-field modeling. Important efforts to validate QSA are similarly limited to the tissue properties and signals that were considered. Especially when new approaches (doses) are developed or novel mechanistic pathways (cellular targets) are considered, the reliance and limits of QSA should be re-considered.
    \item Software packages (numerical solvers) that provide options for “quasistatic” solutions typically reflect the use of this term in the distinct general electromagnetics sense (see section 2.7), and for limited computational purposes. This is not generally sufficient or not necessarily even consistent with neuromodulation QSA as explained here. Rather, regardless of software settings, reports should specify QSA and other modeling assumptions.
    \item Parameters such as assumed tissue conductivity values, boundary conditions (see sections 2.3.2 and 2.3.3), and other model and solver settings should be considered and reported for rigor and reproducibility of the models. The errors due to uncertainties in the parameter values or deviations from any of the assumptions A1--A4 should be estimated if these errors may influence the validity of the conclusions of the study in question. 
\end{enumerate}

Terminology can be varied, ambiguous, and context dependent (e.g., applied electromagnetics versus neuromodulation). 
So irrespective of contemporary nomenclature conventions, the methods in a modeling study should be fully defined to support reproducibility. 
Nonetheless, all papers require some reliance on traditional nomenclature (assumed to be understood across the community). 
We suggest that QSA in neuromodulation papers is understood as adoption of all four assumptions defined herein. 
Referencing the limiting governing equations and fixed tissue conductivities indicates that all four assumptions are applied. 
The relaxation of any of the assumptions is atypical enough in neuromodulation that it should be noted, explicitly and/or through definition of terms (e.g., frequency dependent conductivity).
The precise definition and referencing of QSA in neuromodulation underpin reproducibility and rigor when using QSA models or testing their limits.


\section{Acknowledgments}

Research reported in this publication was supported by the U.S.A. National Institutes of Health under awards No. R01MH128422, R01NS117405, R01NS126376, UG3DA048502, T34 GM137858, R01NS112996, R01NS101362, T32GM136499, and U01NS126052, the European Research Council (ERC) under the European Union’s Horizon 2020 research and innovation programme (grant agreement No. 81037), and the French National Research Agency under the LabEx CominLabs excellence laboratory program (No. ANR-10-LABX-07-01). The content is solely the responsibility of the authors and does not necessarily represent the official views of the funding agencies.

\section{Conflicts of interest}

The authors declare no competing interests relevant to the material presented.

\clearpage
\section*{References}
\bibliographystyle{unsrt}
\bibliography{bibliography}

\end{document}